\documentclass[11pt]{article}
\usepackage{amsmath,amssymb}
\usepackage{blois}

\def\be{\begin{equation}}
\def\ee{\end{equation}}
\def\bea{\begin{eqnarray}}
\def\eea{\end{eqnarray}}

\newcommand{\Photo}

\begin{document}
\vspace*{4cm}
\title{MIXING OF GRAVITONS WITH PHOTONS IN PRIMORDIAL MAGNETIC FIELDS}

\author{ DAMIAN EJLLI }

\address{Dipartimento di Fisica e Scienze Della Terra, Universit\`{a} Degli Studi di Ferrara\\
Via Saragat 1, 44121 Ferrara Italy\\
Astroparticule et Cosmologie, Rue Alice Domon et Leonie Duquet,\\
75205 Paris Cedex 13 France}

\maketitle\abstracts{
Here I discuss the conversion of relic gravitons into photons in large scale cosmological magnetic fields. It is shown that the conversion probability is quite large at the post recombination epoch with a rather large density of formed photons. The produced electromagnetic radiation could make a substantial contribution to the cosmic extragalactic background light and even explain the cosmic x ray background excess.}

\section{Introduction}

During the last years gravitational wave astronomy has been one of the most exciting research area in theoretical and experimental physics. Detection of gravitational waves (GWs) is of highly interest especially in general relativity and would be a further confirmation of its validity. Moreover, their existence is also important in quantum gravity, since its formulations needs the existence of free fields of gravity, namely GWs.
Nowadays there exist several operating GW detectors aiming their detection in the next coming years ($\sim 2016/17$). The most important of them include ground based GW detectors such as Virgo, LIGO, GEO 600 and the planned space interferometers such as eLISA/NGO, DECIGO and BBO.

Most of of the above mentioned detectors operate and will operate in a limited frequency band of the GW spectrum, $f\sim 10^{-4}-10^3$ Hz \cite{gw-rev}. This limited region of the frequency band does not allow to fully explore other parts of the spectrum which might give us important informations about the first seconds after the bing bang. Indeed, GWs from the early Universe would give us important informations about their emission source since they would come to us completely ``untouched'' due to the weak interaction of gravitons. Another reason why to look for GWs from the early Universe is because their energy density is higher than the energy density of GWs emitted by low redshift sources.

In what follows I concentrate on the high frequency part of the spectrum, namely in the frequency range of $f\sim 10^{13}-10^{18}$ Hz. A mechanism of production of such a high frequency background has been proposed in \cite{Dolgov:2011cq} where a population of primordial black holes (PBHs) would emit GWs due to the mechanism of Hawking evaporation. Clearly there are no operating GW detectors in this frequency band and here I consider an alternative way to ``see'' them. This alternative way is based on the mechanism of graviton to photon mixing in an external magnetic field. The inverse process was first proposed in \cite{gertsen} at a classical level and after in  \cite{g-to-gamma} where some laboratory experiments were proposed. In this talk I will review the most important recent results of the graviton to photon mixing in cosmology which is mostly based on the recent papers in  \cite{Dolgov:2012be}.

\section{How do gravitons mix with photons?}

\begin{figure}[htbp]
\begin{center}
\includegraphics[scale=0.5]{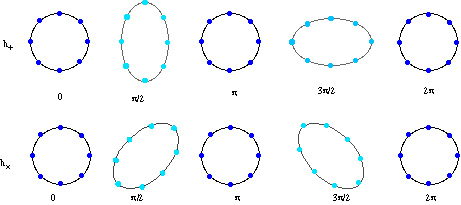}
\caption{The interaction of a GW with an external magnetic field is shown. The field lines are denoted with doted points and its direction is perpendicular to the direction of propagation of the wave. In the upper panel the interaction of a GW with polarization state $+$ and amplitude $h_+$ is shown. In the lower panel the interaction for the polarization state $\times$ with amplitude $h_\times$ is shown. The deformation of the field lines in time create an EMW which propagates in the external field $\mathbf B_e$ creating a time dependent non trivial energy momentum tensor $T_{ij}^{em}$.}
\label{GW-EM}
\end{center}
\end{figure}

It might seems odd, to speak about mixing between the graviton and the photon because of the different spin of the two particles. However, such a mixing is possible in an external magnetic field where the field plays the role of the mixing agent.  In order to see more closely how such a mixing occurs, there two different arguments which one might follow. From a quantum field theoretical approach the external field gives to the photon or graviton the missing quantum number in order for the mixing to occur. Classically the conversion of GWs into electromagnetic waves can be understood by following simple argument. Let us consider a plane GWs which enters into a region where is present a static external magnetic field $\mathbf B_e$. The interactions of the GW with the external field would stretch the field lines and thus its strength would change as $|\mathbf B_e|\rightarrow h(\mathbf x, t)|\mathbf B_e|$ with $h(\mathbf x, t)$ being the GW amplitude. In this case the field would change from static to non static. According to the classical electromagnetism a field which changes in space would produces an electric field, $\nabla\times\mathbf B_e=\partial_t \mathbf E_e$. \footnote{Here, we consider only a static field because it is more simple to deal with it, however one can also consider a field which varies both in space an time but the formalism which follows would be more complicated.} On the other hand the formed electromagnetic wave (EMW) interacts with the background magnetic field and creates a non trivial energy momentum tensor $T_{ij}^{em}\propto B_i^eB_j.$ Since GWs satisfy the wave equation $\Box h_{ij}\propto T_{ij}^\textrm{em}\propto B_i^eB_j$ it implies that the interaction of an EMW with a background field $\mathbf B_e$ creates a GW see, Fig.\ref{GW-EM}.  In this way  it is created an oscillatory phenomena between the GW and the EMW. The efficiency of this kind of oscillation is not $100\%$ as we see in the next sections due to the coherence breaking of photons in the plasma medium.

\begin{figure}[htbp]
\begin{center}
\includegraphics[scale=0.5]{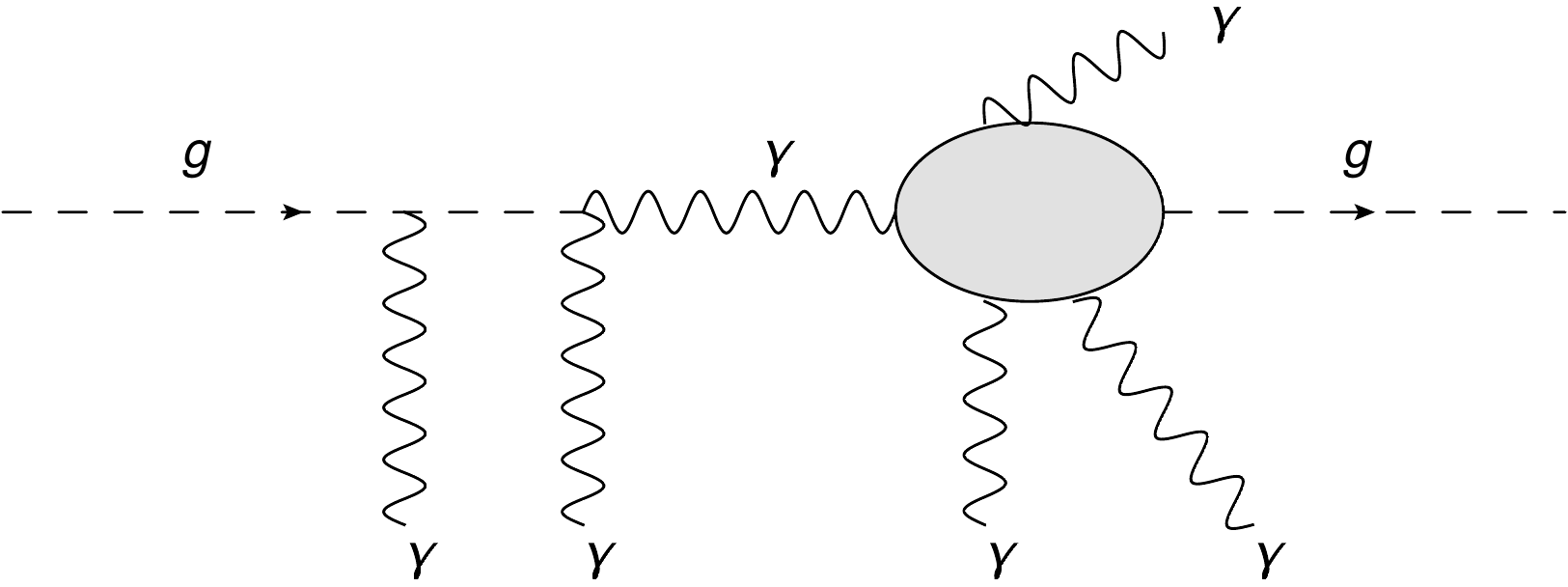}
\caption{Feynman diagram for the oscillation of gravitons into photons in an external background magnetic field. In the initial state we have a graviton $g$ (dashed line) which interacts with two external photons $\gamma$ (wave lines at the bottom) and at the intermediate state we have a photon $\gamma$. After the intermediate photon undergoes to diffraction phenomena due to vacuum polarization effects (black circe of an electron-positron pair) in a magnetic field. The presence of an electron positron pair reduces the photon velocity which gives to the photon a refraction index different from one. The intermediate photon interacts with the external photons producing again a graviton.}
\label{g-gamma mix}
\end{center}
\end{figure}

Having described qualitatively how gravitons transform into photons, now we put the above arguments in convenient formalism. Since the gravitational field interacts with the electromagnetic field, the total Lagrangian density $L$ contains respectively both the free fields kinetic terms and their interaction term, namely
\begin{equation}
L_g=-\frac{1}{4}(\partial_\mu h_{\alpha\beta}\partial^\mu h^{\alpha\beta}-\partial_\mu h\partial^\mu h+2\partial_\mu h^{\mu\nu}\partial_\nu h-2\partial_\mu h^{\mu\nu}\partial_\rho h_\nu^\rho),
\end{equation}
is the gravitational Lagrangian density and
\begin{equation}
L_{{em}}=-\frac{1}{4} F_{\mu\nu}F^{\mu\nu}+\frac{\alpha^2}{90m_e^4}\left[(F_{\mu\nu}F^{\mu\nu})^2+\frac{7}{4}(\tilde F_{\mu\nu}F^{\mu\nu})^2\right],
\end{equation}
 is the Lagrangian density of the electromagnetic field which includes non linear terms due to vacuum polarization effects \cite{EH-Schwinger}. Here we work with the natural units $k_B=\hbar=c=1$ and the metric signature $\eta_{\mu\nu}=(-1, 1, 1, 1)$. The term $\alpha=1/137$ is the fine structure constant, $m_e$ is the electron mass, $F_{\mu\nu}$ is the electromagnetic field tensor and $\tilde F_{\mu\nu}$ is its dual. The interaction of the electromagnetic field with the gravitational field is described by the interaction Lagrangian
 \begin{equation}
L_{I}=\frac{\kappa}{2}h_{\mu\nu}T_{em}^{\mu\nu},
\end{equation}
 where $\kappa=\sqrt{16\pi G}$ is constant which enters in the linear term of the metric expansion $g_{\mu\nu}=\eta_{\mu\nu}+\kappa h_{\mu\nu}$ with $g_{\mu\nu}$ being the total metric tensor. The equation of motion for the total Lagrangian $L=L_g+L_{em}+L_I$ are obtained as usual by the Euler-Lagrange equations of motions. In the limit when the coherence length of the magnetic field $\lambda_B$ is greater than the photon wavelength $\lambda_p$, $\lambda_p\ll \lambda_B$ the equations of motions are well described in the WKB limit \cite{Dolgov:2012be}$^,$\cite{Raffelt:1987im}
 \begin{equation}
\left[(\omega+i\partial_{\mathbf{x}})\mathbf I+
\begin{bmatrix}
  \omega(n-1)_{\lambda} & B_{T}/m_{Pl} \\
  B_{T}/m_{Pl}  & 0  \\
   \end{bmatrix}
\right] 
\begin{bmatrix}
  A_\lambda({\mathbf{x}}) \\
  h_\lambda({\mathbf{x}}) 
 \end{bmatrix}
=0\,,
\label{matrix}
\end{equation}
where $\mathbf I$ is the unit matrix, $m_{Pl}$ is the Planck mass, $\mathbf x$ is the direction of the graviton/photon propagation,
$n$ is the total refraction index of the medium, $\omega$ is the graviton energy, $B_T$ is the strength of the transverse external magnetic field $\mathbf B_e$ and $h_\lambda, A_\lambda$ are  respectively the graviton and photon polarization states with $\lambda$ being the polarization index (helicity) of the graviton and the photon states. In the case of photons $\lambda=+$ indicates a polarization state perpendicular to the external magnetic field and $\lambda=\times$ indicates a state with polarization parallel to the external field. In Fig.\ref{g-gamma mix} the Feynman diagram for the graviton to photon oscillation is shown. The graviton interacting with two external photons changes his angular momentum from two to one and becomes a photon. 

\section{Cosmological implications and results}

The oscillation of gravitons into photons has interesting consequences in cosmology which we are going to discuss. The presence of large scale magnetic fields induces a resonant transition between the gravitons and the photons for graviton energy $\omega\gg m_e$. As a consequence of this transformation, it is produced an almost isotropic background of electromagnetic radiation. The origin of the initial high energy gravitons is assumed to be emitted by a population of PBHs which evaporate before to the bing bang nucleosyntheis epoch, see \cite{Dolgov:2011cq} for more details. The transition probability for such a process can be very high depending on the value of the magnetic field. Here we concentrate only at the post recombination epoch since the conversion probability is very small prior to this epoch. For the values of the magnetic field at present epoch, we use those found in \cite{Paoletti:2012bb}$^,$\cite{Kahniashvili:2008hx}.
\begin{figure}[htbp]
\begin{center}
\includegraphics[scale=0.8]{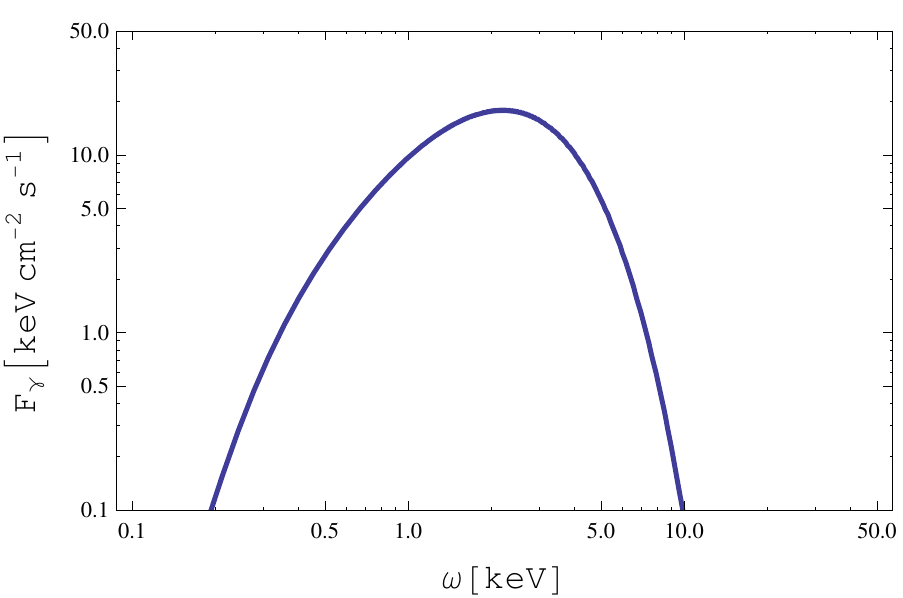}
\caption{Energy flux $F_\gamma$ of photons produced by the transformation of gravitons into photons as a function of the photon energy $\omega$ at the present epoch, for initial value of the magnetic field at recombination $B_i\simeq 2.37$ G.}
\label{Gamma-spectrum}
\end{center}
\end{figure}

However, one needs to take into account coherence breaking of the formed photons in the post recombination plasma. The main effect of damping arises due to Compton scattering of photons into electrons. An important role plays also plasma oscillation in the post recombination epoch which together with QED effects due to vacuum polarization give to the photons a refraction index different from unity. Taking all these effects into account, in Fig.\ref{Gamma-spectrum} the produced electromagnetic energy flux $F_\gamma$ is shown. The flux energy depends essentially on the value of the large scale magnetic field at the post recombination epoch. In the case of the value of the magnetic field at recombination $B_i=2.37$ G which is the redshifted value of the present day magnetic field of those found in \cite{Kahniashvili:2008hx}, the produced spectrum can explain the the excess observed in the cosmic x-ray background (CXB). For smaller value of the magnetic field, the produced electromagnetic radiation would be the main component of the CXB. 

The discussed mechanism is of highly interest in cosmology since it would provide us with an alternative way to indirectly detect GWs in frequency bands where GW detectors do not operate. Moreover, such a mechanism can in principle explain the excess in the CXB and deeper investigation of its spectrum by the present and future electromagnetic detectors would give important hints on this mechanism.

\end{document}